\documentclass[twocolumn]{aastex62}
\shorttitle{Delay time of GW-GRB associations}
\shortauthors{Zhang}

\begin{document}


\title{The Delay Time of Gravitational Wave -- Gamma-Ray Burst Associations}


\author{Bing Zhang}
\affil{Department of Physics and Astronomy, University of Nevada Las Vegas, NV 89154, USA}
\email{zhang@physics.unlv.edu}



\begin{abstract}
The first gravitational wave (GW) - gamma-ray burst (GRB) association, GW170817/GRB 170817A, had an offset in time, with the GRB trigger time delayed by $\sim$1.7 s with respect to the merger time of the GW signal. We generally discuss the astrophysical origin of the delay time, $\Delta t$, of GW-GRB associations within the context of compact binary coalescence (CBC) -- short GRB (sGRB) associations and GW burst -- long GRB (lGRB) associations. In general, the delay time should include three terms, the time to launch a clean (relativistic) jet, $\Delta t_{\rm jet}$; the time for the jet to break out from the surrounding medium, $\Delta t_{\rm bo}$; and the time for the jet to reach the energy dissipation and GRB emission site, $\Delta t_{\rm GRB}$. For CBC-sGRB associations, $\Delta t_{\rm jet}$ and $\Delta t_{\rm bo}$ are correlated, and the final delay can be from 10 ms to a few seconds. For GWB-lGRB associations, $\Delta t_{\rm jet}$ and $\Delta t_{\rm bo}$ are independent. The latter is at least $\sim$10 s, so that $\Delta t$ of these associations is at least this long. For certain jet launching mechanisms of lGRBs, $\Delta t$ can be minutes or even hours long due to the extended engine waiting time to launch a jet. We discuss the cases of GW170817/GRB 170817A and GW150914/GW150914-GBM within this theoretical framework and suggest that the delay times of future GW/GRB associations will shed light into the jet launching mechanisms of GRBs.
\end{abstract}


\keywords{gravitational waves -- gamma-ray bursts: general}



\section{Introduction}

The first neutron star - neutron star (NS-NS) merger gravitational wave source GW170817 \citep{GW170817} was followed by a short gamma-ray burst (GRB) 170817A \citep{GW170817/GRB170817A,goldstein17,zhangbb18b}. The short GRB (sGRB) triggered the Fermi GBM at $\Delta t \sim$ 1.7 s after the merger and lasted for $\sim 2$ s. It consists of two pulses \citep{goldstein17,zhangbb18b}, each lasting for $\sim 1$ s. Earlier, a controversial $\gamma$-ray signal, GW150914-GBM, was claimed by the Fermi GBM team to follow the first black hole - black hole (BH-BH) merger GW event, GW150914, with a delay of $\sim 0.4$ s (\citealt{connaughton16,connaughton18}, cf. \citealt{greiner16}).

The LIGO/Virgo third observational run (O3) started on April 1, 2019 and will last for one year. It is highly expected that more GW-GRB associations will be detected. At least NS-NS mergers and NS-BH mergers with a mass ratio $q \equiv m_1 / m_2 < 5$ ($m_1 > m_2$) are expected to produce  sGRBs\footnote{NS-BH mergers with $q > 5$ would not produce a GRB since the NS would not be tidally disrupted before being swallowed by the BH as a whole \citep[e.g.][]{shibata09}.}. If some BH-BH mergers can make GRBs, more robust cases than GW150914/GW150914-GBM should be identified. Finally, under certain conditions, core collapse events that make long GRBs (lGRBs) may have a strong enough GW signal to be detected as a GW burst (GWB) by LIGO/Virgo GW detectors \citep[e.g.][]{kobayashimeszaros03}.  It is possible that GWB-lGRB associations may be detected in the future.

The delay time of a GRB with respect to the GW signal can not only be used to constrain fundamental physics \citep[e.g.][]{wei17,shoemaker18}, but  also carries important information about GRB physics, including jet launching mechanism, jet breakout from the surrounding medium, jet dissipation, and GRB radiation mechanism. All these are closely related to the unknown composition of the GRB jet, which is subject to intense debate in the field of GRBs (\citealt{zhang18} for a comprehensive discussion). The origin of the $\sim 1.7$ s delay in GW170817/GRB 170817A has been discussed in the literature \citep[e.g.][]{granot17,zhangbb18b,shoemaker18,veres18,lin18,salafia18}. In this mini-review, we systematically investigate several physical processes that contribute to the observed time delay $\Delta t$  (Section \ref{sec:delay}). This is discussed within the context of compact binary coalescence (CBC)-sGRB associations and GWB-lGRB associations (Section \ref{sec:short-long}). The case studies for GW170817/GRB 170817A and GW150914/GRB 150914-GBM are presented in Section \ref{sec:cases}. The results are summarized in Section \ref{sec:summary} with some discussion.

\section{Delay time of GW-GRB associations}\label{sec:delay}

\subsection{General consideration}

We assume that both GWs and photons travel with speed of light and only discuss the astrophysical origin of $\Delta t$. For discussions on how to use $\Delta t$ to constrain physics beyond the standard model, see \cite{wei17}, \cite{shoemaker18} and references therein.

\begin{figure}
\plotone{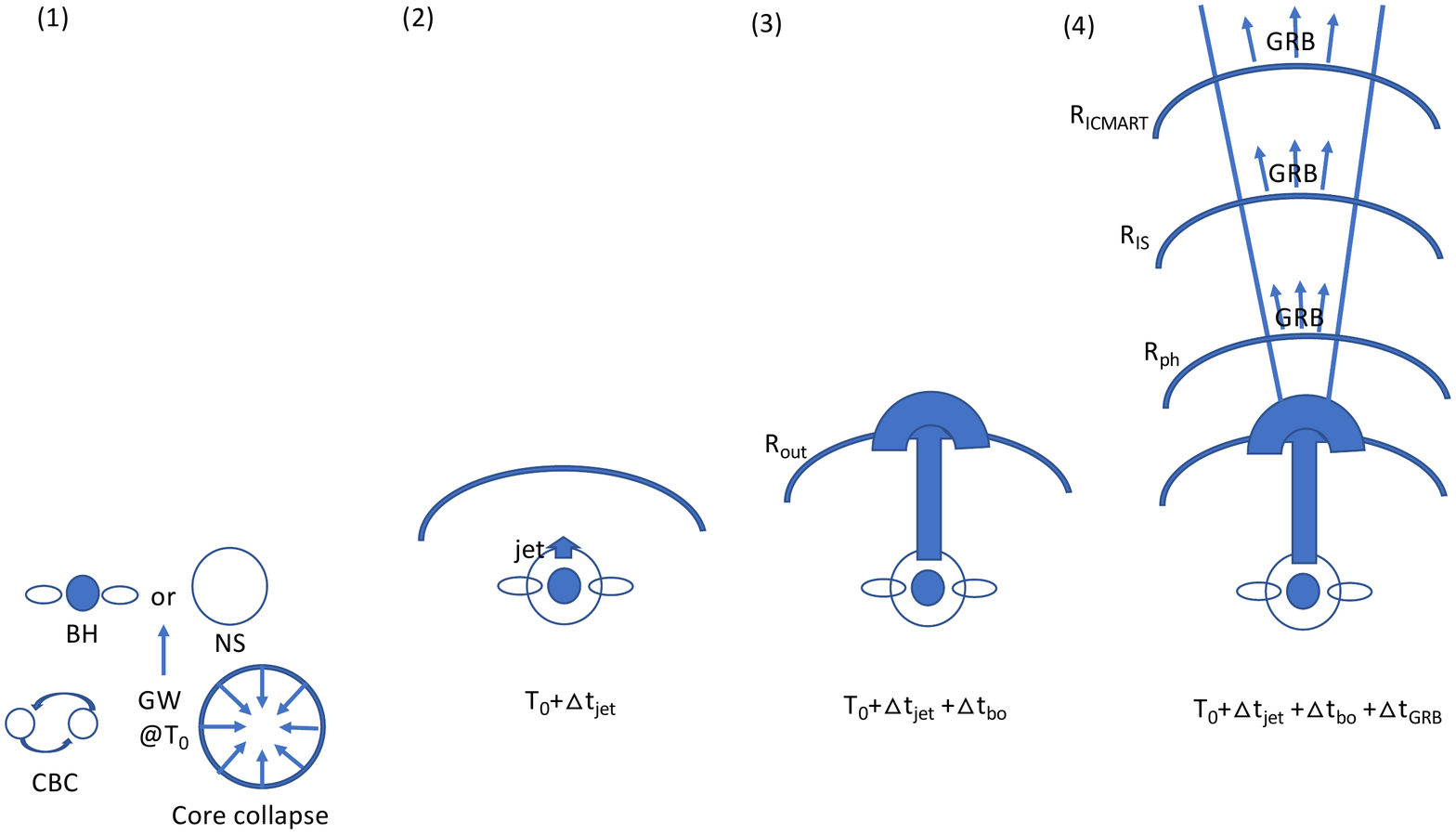}
\caption{A cartoon picture of a generic GW-GRB association. There are four steps: (1) A CBC or a core collapse event makes a bright GW signal, either in the form of a ``chirp'' or a ``GWB''. This marks the zero time point $T_0$. The event produces a hyper-accreting BH or a rapidly spinning magnetar; (2) After $\Delta t_{\rm jet}$, a clean jet is launched from the central engine; (3) After another time interval $\Delta t_{\rm bo}$, the jet breaks out from the surrounding medium; (4) After another time interval $\Delta t_{\rm GRB}$, the jet reaches the GRB radius where $\gamma$-rays are emitted. Depending on the jet composition, there could be three possible sites: $R_{\rm ph}$, $R_{\rm IS}$, and $R_{\rm ICMART}$, which correspond to three different durations of $\Delta t_{\rm GRB}$. 
\label{fig1}}
\end{figure}

In general, the observed delay time due to an astrophysical origin should consist of three terms (Fig. \ref{fig1}), i.e.
\begin{equation}
 \Delta t = (\Delta t_{\rm jet} + \Delta t_{\rm bo} + \Delta t_{\rm GRB}) (1+z),
\label{eq:Deltat}
\end{equation}
where $\Delta t_{\rm jet}$ is the time for the engine to launch a relativistic jet, $\Delta t_{\rm bo}$ is the time for the jet to penetrate through and break out from the surrounding medium (the ejecta in the CBC scenario and the stellar envelope in the core collapse scenario), and $\Delta t_{\rm GRB}$ is the time after breakout for the jet to reach the energy dissipation radius where the observed $\gamma$-rays are emitted. The $(1+z)$ factor is the cosmological time dilation factor, which we will ignore in the rest of the discussion. All three time intervals are measured in the rest frame of the Earth observer (with the $(1+z)$ correction). Since the engine is at rest with respect to the observer (ignoring proper motion of both the source and Earth) and since the jet is propagating with a non-relativistic speed in the surrounding medium, $\Delta t_{\rm jet}$ and $\Delta t_{\rm bo}$ are also essentially the times $\Delta \hat t_{\rm jet}$ and $\Delta \hat t_{\rm bo}$ measured in the rest frame of the central engine, which we call the ``lab frame''.

\subsection{$\Delta t_{\rm jet}$}
The jet launching time $\Delta t_{\rm jet}$ depends on the type of the central engine and the jet launching mechanism of GRBs. In the literature, a GRB jet can be launched either through accretion (onto a BH or a NS) or a magnetic mechanism. The latter applies to an NS (magnetar) engine. A relativistic jet is launched either as magnetic bubbles generated from differential rotation of the NS or through magnetic dipole spindown of a rapidly spinning magnetar. 

For an accreting central engine, $\Delta t_{\rm jet}$ can be decomposed into three terms: 
\begin{equation}
\Delta t_{\rm jet,acc} = \Delta t_{\rm wait} + \Delta t_{\rm acc} + \Delta t_{\rm clean},
\end{equation}
where $\Delta t_{\rm wait}$ is the waiting time for a specific accretion model to operate; $\Delta t_{\rm acc}$ is the timescale to form the accretion disk and to start accretion; and $\Delta t_{\rm clean}$ is the timescale to launch a relativistic jet since accretion starts, which requires that mass loading is low enough, i.e. the mass loading parameter
\begin{equation}
\mu_0 \equiv \eta \Gamma_0 (1+\sigma_0) \equiv L_{\rm jet}(t)/\dot M_{\rm jet}(t) c^2 \gg 1,
\end{equation} 
where $\eta$ is the dimensionless enthalpy of the jet at the central engine, $\Gamma_0 \sim 1$ is the initial Lorentz factor of the jet, and $\sigma_0$ is the ratio between the magnetic energy density and matter energy density (including internal energy) at the central engine. Here $L_{\rm jet}(t)$ is the time-dependent jet luminosity, and $\dot M_{\rm jet}(t)$ is the time-dependent mass loading rate in the jet. 

For the magnetic engine model, the jet launching time can be decomposed as 
\begin{equation}
\Delta t_{\rm jet,mag} =  \Delta t_{\rm wait} +  \Delta t_{\rm B} + \Delta t_{\rm clean},
\end{equation} 
where $\Delta t_{\rm wait}$ is again the waiting time for a specific magnetic model to operate, $\Delta t_{\rm B}$ is the timescale to establish a strong magnetic field through differential rotation, and $\Delta t_{\rm clean}$ is again the timescale for the environment to become clean enough to launch a relativistic jet. 

A few more words about $\Delta t_{\rm wait}$: Since the GRB jet launching mechanism is not identified, different jet launching models make different assumptions. When we discuss a particular model, $\Delta t_{\rm wait}$ is defined as the waiting time when the conditions for that mechanism to operate are satisfied. For example, for a model invoking a hyper-accreting BH to launch a jet, $\Delta t_{\rm wait}$ is the waiting time for a BH to form, which is the lifetime of a hypermassive neutron star (HMNS) or a supramassive neutron star (SMNS) before collapsing. Within this model, no jet is launched if the central object is an NS. On the other hand, under the same physical condition but for the jet launching model invoking a differentially rotating NS, a jet is directly launched during the HMNS phase, so that $\Delta t_{\rm wait}=0$. For another example, in the magnetar model, if the jet launching mechanism is through magnetic spindown, the early brief accretion phase may be regarded as $\Delta t_{\rm wait}$. On the other hand, in the model invoking a hyper-accreting NS, $\Delta t_{\rm wait}=0$ for the same physical condition. See Section \ref{sec:short-long} for a detailed discussion on different jet-launching models.

Several timescales are related to the dynamical timescale of the system
\begin{equation}
 t_{\rm dyn} = 2\pi \left({R^3 \over GM}\right)^{1/2} \simeq 1.8 \ {\rm ms} \ R_{6.5}^{3/2}
 \left(\frac{M}{3 {\rm M_\odot}}\right)^{-1/2},
 \label{eq:tdyn}
\end{equation}
where $M$ and $R$ are the mass (normalized to $3 \rm M_\odot$, where $\rm M_\odot$ is the solar mass) and radius of the central engine (convention $Q_n = Q/10^n$ in cgs units adopted throughout the paper). In the accretion model, the disk forms within $t_{\rm dyn}$, and the accretion starts within the viscous timescale $\sim \alpha^{-1} t_{\rm dyn} = 10 \alpha_{-1}^{-1} t_{\rm dyn}$, where $\alpha \sim 0.1$ is the dimensionless viscosity constant. As a result, $\Delta t_{\rm acc} \sim n t_{\rm dyn}$, where $n \sim 10$. Similarly, in the magnetar model, magnetic amplification also takes a few dynamical timescales, i.e. $\Delta t_{\rm B} \sim n t_{\rm dyn}$, where $n \sim $ a few.  

In both models, the timescale for the jet to become clean, $\Delta t_{\rm clean}$ is defined by the degree of mass loading. For a new born, hot central engine (either the hot accretion disk or the central magnetar), the dominant mass-loading mechanism is through the neutrino wind, with the mass-loading rate of an unmagnetized flow defined by \citep{qian96}
\begin{equation}
 \dot M_\nu = (2.5\times 10^{-5} \ {\rm M_\odot/s}) L_{\nu,52}^{5/3} \left(\frac{\epsilon_\nu}{10 \ {\rm MeV}}\right)^{10/3} R_6^{5/3} \left(\frac{M}{3\rm M_\odot}\right)^{-2},
\end{equation}
where $L_\nu$ is the $\nu / \bar\nu$ luminosity, and $\epsilon_\nu$ is the typical energy of $\nu/\bar\nu$. A magnetized engine will suppress mass loading by limiting entry of protons into the jet \citep{lei13}. A detailed treatment of mass loading for the black hole and magnetar central engines have been carried out by \cite{lei13} and \cite{metzger11}, respectively.

\subsection{$\Delta t_{\rm bo}$}

After a clean jet with $\mu_0 \gg 1$ is launched, it has to propagate through the dense medium surrounding the engine. For the case of CBC-sGRB associations, the surrounding medium is mostly the ejecta launched right before the merger. For the case of GWB-lGRB associations, the surrounding medium is the in-falling stellar envelope of the progenitor star.

{In order to launch a successful jet, a critical value of the jet energy needs to be reached \citep[e.g.][]{Duffell18}. In the following, we assume that such a condition is satisfied, which is justified by the observations of GRB 170817A that show evidence of a successful jet \citep{mooley18b,ghirlanda19}. }
Very generally, the jet breakout timescale can be written as
\begin{equation}
\Delta t_{\rm bo} = \frac{R_{\rm out} - R_{\rm in}}{(\beta_{\rm jet,h} - \beta_{\rm out})c},
\label{eq:t_bo}
\end{equation}
where $R_{\rm out}$ and $\beta_{\rm out}$ are the radius and dimensionless velocity (in the rest frame of the central engine) of the outer boundary of the surrounding medium, $R_{\rm in}$ is the radius of the central engine where the jet is launched, and $\beta_{\rm jet,h}$ is the dimensionless speed of the jet head propagating inside the medium. Notice that $\beta_{\rm jet,h}$ is much smaller than the termination dimensionless speed of the jet, $\beta_{\rm jet}$, due to the high density of the surrounding medium. Whereas $\beta_{\rm jet} \sim 1$ for a relativistic jet, $\beta_{\rm jet,h} = 0.1 \beta_{\rm jet,h,-1}$ is typically non-relativistic. 

For the case of CBC-sGRB associations, the dynamical ejecta moves outward with a dimensionless speed $\beta_{\rm out} = \beta_{\rm ej} = 0.1 \beta_{\rm ej,-1}$. The outer boundary of the ejecta\footnote{Note that the ejecta has a velocity profile, with the outer boundary defined by the fastest layer in the ejecta. For an order-of-magnitude treatment, we adopt the average speed of the ejecta.} is defined as $R_{\rm out} = R_{\rm ej} = \beta_{\rm ej} c (\Delta t_{\rm jet} + \Delta t_{\rm tidal})$, where $\Delta t_{\rm tidal}$ is the time interval between the epoch when the neutron star is tidally disrupted (i.e. the dynamical ejecta is launched) and the epoch of coalescence. This is typically of the order of milliseconds \citep{shibata08}.  

In order to break out the ejecta, the jet head needs to propagate faster than the ejecta. Let the jet head advance with a dimensionless speed $\beta'_{\rm jet,h}$ in the ejecta frame. Its lab-frame dimensionless speed reads
\begin{equation}
\beta_{\rm jet,h} = \frac{\beta'_{\rm jet,h} + \beta_{\rm ej}}{1+\beta'_{\rm jet,h} \beta_{\rm ej}}.
\end{equation}
Equation (\ref{eq:t_bo}) can be then written as
\begin{equation}
\Delta t_{\rm bo} ({\rm CBC/sGRB}) \simeq 
\frac{\beta_{\rm ej} (\Delta t_{\rm jet} + \Delta t_{\rm tidal}) - R_{\rm in}/c}{\beta_{\rm jet,h} - \beta_{\rm ej}} ,
\end{equation}
When $\Delta t_{\rm jet} \gg \Delta t_{\rm tidal}$, one has $R_{\rm out} \gg R_{\rm in}$, so that $\Delta t_{\rm bo} \simeq \frac{\beta_{\rm ej}}{\beta_{\rm jet,h}-\beta_{\rm ej}} \Delta t_{\rm jet} \simeq (\beta_{\rm ej} / \beta'_{\rm jet,h}) \Delta t_{\rm jet} \propto \Delta t_{\rm jet}$ (see also \cite{geng19}). In this case, the first two terms of Eq.(\ref{eq:Deltat}) are correlated with each other. {Since different bursts likely have different $\beta_{\rm ej}$ and $\beta'_{\rm jet,h}$, the positive correlation between $\Delta t_{\rm jet}$ and $\Delta t_{\rm bo}$ should have a broad scatter.}

For the case of GWB-lGRB associations, $R_{\rm out}$ is the outer boundary of the stellar envelope, $R_*$, which may be regarded as a constant during jet propagation, even if it is slowly shrinking due to fallback. Since $R_* \gg R_{\rm in}$, the breakout time in this case is 
\begin{equation}
\Delta t_{\rm bo} ({\rm GWB/lGRB}) \simeq \frac{R_{\rm *}}{\beta_{\rm jet,h}c}.
\label{Deltatbo2}
\end{equation}

\subsection{$\Delta t_{\rm GRB}$}

Since the jet travels with a relativistic speed after breaking out the surrounding medium (the brief acceleration phase ignored), the observer-frame time $\Delta t_{\rm GRB} = (1-\beta \cos\theta) \Delta \hat t_{\rm GRB}$, where $\Delta \hat t_{\rm GRB}$ is the lab-frame duration for the jet to travel to the GRB emission radius, $\beta$ is the Lorentz factor and dimensionless speed of the relativistic jet, and $\theta$ is the angle between the jet direction and the line of sight. Since $\Delta \hat t_{\rm GRB} \simeq R_{\rm GRB} / c$, one has
\begin{equation}
\Delta t_{\rm GRB} \simeq (1 - \beta \cos\theta) \frac{R_{\rm GRB}}{c} \simeq  \frac{R_{\rm GRB}}{\Gamma^2 c}.
\label{eq:t_GRB}
\end{equation}
The last approximation is valid when $\theta \sim (0-1/\Gamma)$ with $\Gamma$ being the Lorentz factor along the line of sight, which is relevant for a relativistically moving outflow with the $1/\Gamma$ cone covering the line-of-sight. For GW170817/GRB 170817A, simple arguments have ruled out the scenario invoking a top-hat jet beaming away from the line of sight \citep[e.g.][]{granot17,zhangbb18b,ioka18}.

The GRB emission radius is not identified. In the literature, there are at least three sites that have been suggested to emit $\gamma$-rays, which are:
\begin{itemize}
\item The photosphere radius \citep[e.g.][]{meszarosrees00,rees05,peer11}
\begin{equation}
 R_{ph} \simeq (6 \times 10^{12} \ {\rm cm}) L_{w,52} \Gamma_2^{-3},
\end{equation}
where $L_w$ is the isotropic-equivalent ``wind'' luminosity in the line-of-sight direction. Here the low-enthalpy-regime ($\eta < \eta_*$ in the notation of \cite{meszarosrees00}) has been adopted, which is relevant for weak GRBs associated with CBCs at a large viewing angle, such as GW170817/GRB 170817A.
\item The internal shock radius \citep{rees94,kobayashi97}
\begin{equation}
 R_{\rm IS} = \Gamma^2 c \delta t \simeq (3\times 10^{12} \ {\rm cm}) \Gamma_2^2 \delta t_{-2},
\end{equation}
where $\delta t$ is the variability timescale in the lightcurve, which is typically 10s of milliseconds.
\item The internal collision-induced magnetic reconnection and turbulence (ICMART) radius \citep{[e.g.][]zhangyan11,uhm16b}
\begin{equation}
 R_{\rm ICMART} = \Gamma^2 c t_{\rm pulse} \simeq (3\times 10^{14} \ {\rm cm}) \Gamma_2^2 t_{\rm pulse},
\end{equation}
where $t_{\rm pulse}$ is the duration of the broad pulses in the GRB lightcurve, which is typically seconds.
\end{itemize}
Which radius is relevant for GRB emission depends on the composition of the jet. For a matter dominated fireball, a quasi-thermal emission from the photosphere and a synchrotron emission component from the internal shock are expected \citep{meszarosrees00,daigne02,peer06,zhangpeer09}. For a Poynting flux dominated outflow, both the photosphere and internal shock emission components are suppressed, and the GRB emission site is at a large radius $R_{\rm ICMART}$ \citep{zhangyan11}.

Putting three cases together, one has
\begin{eqnarray}
 \Delta t_{\rm GRB} & = & \left\{
  \begin{array}{ll}
    (200 \ {\rm s}) (1-\beta\cos\theta) L_{w,52} \Gamma_2^{-3}, & R_{\rm GRB} = R_{\rm ph}, \\
    (1-\beta\cos\theta)\Gamma^2 \delta t, & R_{\rm GRB} = R_{\rm IS}, \\
   (1-\beta\cos\theta)\Gamma^2 t_{\rm pulse}, & R_{\rm GRB} = R_{\rm ICMART}
  \end{array}
  \right.
 \nonumber \\
& \simeq & \left\{
  \begin{array}{ll}
    (20 \ {\rm ms}) L_{w,52} \Gamma_2^{-5}, & R_{\rm GRB} = R_{\rm ph}, \\
    \delta t, & R_{\rm GRB} = R_{\rm IS}, \\
    t_{\rm pulse}, & R_{\rm GRB} = R_{\rm ICMART},
  \end{array}
  \right.
\label{eq:DeltatGRB}
\end{eqnarray}
where the second part of the equation makes the assumption that there is relativistic moving materials along the line of sight.

\subsection{Burst duration $T_{\rm burst}$}

It is relevant to discuss the true duration, $T_{\rm burst}$, of a GRB here. Note that observationally defined duration $T_{90}$ is the lower limit of $T_{\rm burst}$, since it is limited by the detector's sensitivity. 

GRBs usually show highly variable lightcurves, sometimes displaying multiple broad ``pulses'' with rapidly varying spikes superposed on top \citep{gao12}. Some GRBs only have one or two broad pulses. For GRBs with multiple pulses, the total duration $T_{\rm burst}$ is defined by the duration of the central engine activity. The duration of a broad pulse, $t_{\rm pulse}$, has two different interpretations. For models that invoke a small emission radius ($R_{ph}$ or $R_{\rm IS}$), one has $R_{\rm GRB}/\Gamma^2 c \ll t_{\rm pulse}$, so $t_{\rm pulse}$ has to be interpreted as the duration of one episode of the central engine activity. The broad pulses may be attributed to the modulation at the central engine, e.g. the interaction between the jet and the stellar envelope in the long GRB model \citep[e.g.][]{morsony10}. The hard-to-soft evolution of the peak energy $E_p$ across broad pulses posed a challenge to such an interpretation \citep{deng14b}. Within the ICMART model \citep{zhangyan11} or the large-radius internal shock model \citep{bosnjak14b}, $t_{\rm pulse}$ is interpreted as emission duration of one fluid shell as it expands in space, with the peak time defined as either the time when the shell reaches the maximum dissipation, or when the synchrotron spectral peak sweeps across the observational band \citep{uhm16b,uhm18}. For bursts with one single broad pulse, the burst duration is defined by
\begin{equation}
 T_{\rm burst} = t_{\rm pulse} \simeq \frac{R_{\rm GRB}}{\Gamma^2 c},
\end{equation}
which is the same expression as $\Delta t_{\rm GRB}$ (Eq.(\ref{eq:t_GRB})).

\section{Different GW-GRB association models}\label{sec:short-long}

With the above preparation, in the following we discuss the delay times for different GW-GRB association systems in different models. The results are summarized in Table \ref{tab:Deltat}.

\begin{deluxetable*}{ccccccccc}
\tablecaption{$\Delta t$ in different GW-GRB association scenarios \label{tab:Deltat}}
\tablewidth{0pt}
\tablehead{
\colhead{System} & \colhead{Engine} & \colhead{Jet mechanism} & \colhead{} &
\colhead{$\Delta t_{\rm jet}$} & \colhead{} & \colhead{$\Delta t_{\rm bo}$} & \colhead{$\Delta t_{\rm GRB}$} & \colhead{$\Delta t$} \\
\hline
\colhead{} & \colhead{} & \colhead{} & \colhead{$\Delta t_{\rm wait}$} &
\colhead{$\Delta t_{\rm acc}/\Delta t_{\rm B}$} & \colhead{$\Delta t_{\rm clean}$} & \colhead{} & \colhead{} & \colhead{} 
}
\startdata
BH-NS & BH & accretion & $\sim$ 0 s & $\sim 10$ ms & $\sim$ 0 s & (10-100) ms & $<$ ms to $\sim$ s & (0.01-few) s \\
NS-NS & BH & accretion & $\sim$ 0 s & $\sim 10$ ms & $\sim$ 0 s & (10-100) ms & $<$ ms to $\sim$ s & (0.01-few) s  \\
NS-NS & HMNS/BH & accretion & (0.1-1) s & $\sim 10$ ms & $\sim$ 0 s & (0.1-1) s & $<$ ms to $\sim$ s & (0.1-few) s   \\
NS-NS & HMNS/BH & magnetic &  $\sim$ 0 s & $\sim 10$ ms & (0-1) s & (0.01-1) s & $<$ ms to $\sim$ s & (0.01-few) s  \\
NS-NS & SMNS/SNS & accretion & $\sim$ 0 s & $\sim 10$ ms & (0-0.1) s & (10-100) ms & $<$ ms to $\sim$ s & (0.01-few) s  \\
NS-NS & SMNS/SNS & magnetic & $\sim$ 0 s & $\sim 10$ ms &  (0-10) s  & (0.01-10) s  & $<$ ms to $\sim$ s & (0.01-10) s \\
\hline
Type I collapsar & BH & accretion & (0-several) s & $\sim 10$ ms  & $\sim$ 0 s & (10-50) s  & $<$ ms to $\sim$ s & (10-50) s \\
Type II collapsar & BH & accretion & ($10^2$-$10^4$) s & $\sim 10$ ms & $\sim$ 0 s & (10-50) s  & $<$ ms to $\sim$ s & ($10^2$-$10^4$) s \\
core collapse  & magnetar & accretion & $\sim$ 0 s & $\sim 10$ ms & (0-10) s & (10-50) s  & $<$ ms to $\sim$ s & (10-50) s \\
core collapse  & magnetar & magnetic & $\sim$ 0 s & $\sim 10$ ms & (0-10) s & (10-50) s  & $<$ ms to $\sim$ s & (10-50) s \\
core collapse  & magnetar & spindown & (1-$10^3$) s & N/A & $\sim$ 10 s  & (10-50) s  & $<$ ms to $\sim$ s & (10-$10^3$) s \\
\enddata
\end{deluxetable*}

\subsection{CBC-sGRB associations}

BH-NS mergers with a moderate mass ratio ($q < 5$, \citealt{shibata09}) are expected to be associated with sGRBs. The central engine is a BH, and the jet launching mechanism is accretion. One has $\Delta t_{\rm wait}=0$ s. The dynamical timescale (Eq.(\ref{eq:tdyn})) is $\sim$ ms, so $\Delta t_{\rm acc}$ is $\sim 10$ ms. A hyper-accreting black hole engine is considered clean, especially if energy is tapped via a global magnetic field \citep{lei13}. One can take $\Delta t_{\rm clean} \sim 0$ s. Overall, one would expect $\Delta t_{\rm jet} \sim $10 ms. Assuming $\beta'_{\rm jet,h} \sim 0.1$, $\beta_{\rm ej} \sim 0.1$, and $R_{\rm in} \sim $ 3 times of  the Schwarzschild radius, one gets $\Delta t_{\rm bo} \sim \Delta t_{\rm jet} + \Delta t_{\rm tidal}$, which is also $\sim 10$ ms. Finally, $\Delta t_{\rm GRB}$ depends on jet composition and dissipation mechanism (Eq.(\ref{eq:DeltatGRB})). For a $\nu\bar\nu$-annihilation driven fireball, $\Delta t_{\rm GRB}$ is usually very short ($<$ ms) (unless $\Gamma$ is extremely low) thanks to the small emission radius. For a Poynting-flux-dominated jet, $\Delta t_{\rm GRB} \sim t_{\rm pulse}$ can be up to the duration of the sGRB itself.

NS-NS mergers are more complicated. Depending on the equation of state and the total mass in the merger, there could be four different outcomes \citep[e.g.][]{baiotti17}: a promptly formed BH, a differential-rotation-supported hypermassivs NS (HMNS) followed by collapse, a uniform-rotation-supported supramassive NS (SMNS) followed by collapse, or a stable NS (SNS). The prompt BH case is similar to the BH-NS merger case. For the HMNS case (which forms a BH within $\sim$ (0.1-1) s after the merger) and the SMNS/SNS case (which forms a long-lived NS), we consider both an accretion-powered jet and a magnetically powered jet, respectively. 

The HMNS/BH accretion model assumes that the jet is launched when the BH is formed. This model introduces a waiting time $\Delta t_{\rm wait} \sim \Delta t_{\rm HMNS} \sim (0.1-1)$ s, where $\Delta t_{\rm HMNS}$ is the lifetime of the HMNS. The jet launching time $\Delta t_{\rm jet}$ is dominated by $\Delta t_{\rm wait}$, and the jet breakout time $\Delta t_{\rm bo}$ scales up with $\Delta t_{\rm jet}$ proportionally. The third time $\Delta t_{\rm GRB}$ again can be from shorter than ms (fireball photosphere emission) to the duration of the burst itself (Poynting jet). 

The HMNS/BH magnetic model assumes that a jet can be launched during the HMNS phase, so that $\Delta t_{\rm wait} = 0$ s. The magnetic field amplification time $\Delta t_{\rm B}$ is again several times of $t_{\rm dyn}$, i.e. $\sim 10$ ms. The ``clean'' time $t_{\rm clean}$ is quite uncertain. \cite{rosswog03} performed magnetohydrodynamic (MHD) simulations of NS-NS mergers in the HMNS phase and claimed that magnetic fields can be amplified to several times of $10^{17}$ G within $\Delta t_{\rm B} \sim$ 10 ms. They suggested that a relativistic short GRB jet can be launched along the axis of the binary orbit during this period of time, even though their numerical simulations were not able to resolve the baryon loading process. Within this scenario, $\Delta t_{\rm clean}$ may be as short as 0 s, so that $\Delta t_{\rm jet}$ is of the order of $\Delta t_{\rm B} \sim$ 10 ms. The breakout time $\Delta t_{\rm bo}$ is correspondingly short. On the other hand, a recent numerical simulation \citep{ciolfi19} suggested that no relativistic jet is launched before 100 ms due to strong baryon loading. If this is the case, $\Delta t_{\rm clean}$ may be at least $\sim 0.1$ s for this model. Both $\Delta t_{\rm jet}$ and $\Delta t_{\rm bo}$ are increased correspondingly. Again $\Delta t_{\rm GRB}$ is negligible in the fireball model and can be of the order of the burst duration in the Poynting jet model.

Finally, the SMNS/SNS models assume that such systems can power sGRBs \citep{dai06,gaofan06,metzger08}. The observational evidence of this class of models is the so-called ``internal X-ray plateau'' (a lightcurve plateau followed by a sharp drop of flux which can be only interpreted as ``internal'' dissipation of a central engine outflow) observed following a good fraction of sGRBs \citep{rowlinson10,rowlinson13,lv15}. The plateau is best interpreted as internal dissipation of a post-merger massive magnetar wind \citep{zhang13,gao16a,sun17,xue19} during the magnetic spindown phase \citep{zhangmeszaros01a}. The sGRB needs to be produced by the massive NS, likely shortly after the merger. 

The SMNS/SNS accretion model assumes that a relativistic sGRB jet can be launched from a magnetar via hyper-accretion \citep{metzger08,zhangdai10}. Within this scenario, $\Delta t_{\rm wait} = 0$ s, and accretion starts after $\Delta t_{\rm acc} \sim$ 10 ms. Baryon loading in this model has not been well-studied, and it is assumed that $\Delta t_{\rm clean}$ is short, e.g. $0-100$ ms. As a result, both $\Delta t_{\rm jet}$ and $\Delta t_{\rm bo}$ are tens of ms. $\Delta t_{\rm GRB}$ is again from $<$ ms (fireball) to $\sim 1$ s (Poynting jet).

The SMNS/SNS magnetic model is similar to the HMNS/BH magnetic model, except that $\Delta t_{\rm clean}$ can be longer (no longer limited by $\Delta t_{\rm HMNS}$). Within the most optimistic model suggested by \cite{rosswog03}, $\Delta t_{\rm clean}$ can be as short as $\sim$ 0 s. On the other hand, according to the calculation of \cite{metzger11}, baryon mass loss rate in a proto-NS could be initially very high, e.g. $\sigma_0 <1$ before $\sim 2$ s and $\sigma_0 <100$ before $\sim 10$ s. Within this scenario, the outflow is initially non-relativistic, and a clean jet capable of producing a sGRB is launched only after $\Delta t_{\rm clean} \sim (1-10)$ s. A longer $\Delta t_{\rm clean}$ leads to longer $\Delta t_{\rm jet}$ and $\Delta t_{\rm bo}$, which can exceed $\Delta t_{\rm GRB}$ even for the Poynting jet case. The total delay time $\Delta t$ could be dominated by $\Delta t_{\rm clean}$.

\subsection{GWB-lGRB associations}

Within the core collapse model of long GRBs, the violent collapsing process may leave behind a central object with a significant quadruple moment to generate a GW burst \citep[e.g.][]{kobayashimeszaros03}. For these systems, unlike CBCs, it is not straightforward to define the epoch of significant GW radiation (i.e. the peak of GW burst signal). The strongest GW radiation is likely produced during the core collapse phase \citep[e.g.][]{ott09}. The rapidly proto-NS (magnetar) may carry a significant quadruple moment and radiate GW emission as well \citep{usov92,zhangmeszaros01a,corsi09}. Finally, after the NS collapses to a BH, the neutrino-dominated accretion flow (NDAF) into the BH may also radiate GW, even if with a lower amplitude \citep[e.g.][]{liu17b}. In the following discussion, we assume that the GWB emission peaks at the core collapse time.

One can consider two general categories of models: core collapse events leading to a BH engine (which is usually called the ``collapsar'' model), and core collapse events leading to a magnetar engine (which we call the ``magnetar'' model). The progenitor of both types of engines can be either an isolated single star or a binary system whose merger induces the core collapse of the merged star. 

Two types of collapsars have been discussed in the literature: Type I collapsar model  \citep{woosley93,macfadyen99} invokes the collapse of the iron core of a rapidly rotating helium star, forming a short-lived NS that subsequently collapses in a few seconds. Type II collapsar model \citep{macfadyen01}, on the other hand, invokes a long-lived NS, which continues to accrete fall-back materials for an extended period of time before collapsing to a BH minutes or even hours later. The progenitor stars of Type II collapsars are more common, so that these events may have a higher rate than Type I collapsars \citep{macfadyen01}. For either case, since a BH is required to launch a GRB jet, there is a waiting time $\Delta t_{\rm wait}$ that marks the duration of the proto-NS phase, ranging from several seconds (Type I collapsar) to hours (Type II collapsar). This term is likely the dominating term in $\Delta t_{\rm jet}$. Unlike CBC-sGRB associations, the jet breakout time $\Delta t_{\rm bo}$ is independent of $\Delta t_{\rm jet}$ when $\Delta t_{\rm jet}$ is smaller than the free-fall timescale of a massive star $t_{\rm ff} = [3 \pi/(32 G \bar\rho)]^{1/2} \sim 180 \ {\rm s} \ (\bar\rho/100 \ {\rm g \cdot cm^{-3}})^{-1/2}$ ($\bar\rho$ is the mean density of the stellar envelope),  and is set by the size of the progenitor and the jet head speed. The widely accepted progenitor system of lGRBs is Wolf-Rayet stars \citep{woosley06}. Taking $R_* \sim 10^{11}$ cm and $\beta_{\rm jet,h} \sim 0.1$, one gives $\Delta t_{\rm bo} \sim (10-50)$ s based on Eq.(\ref{Deltatbo2}), which is longer than $\Delta t_{\rm GRB}$. The final $\Delta t$ is $\sim$ tens of seconds for Type I collapsar and $\sim (10^2 - 10^4)$ s for Type II collapsar. 

For the magnetar model, a long GRB may be produced via one of the following three mechanisms: accretion, magnetic due to differential rotation, and magnetic due to spindown. The three mechanisms differ mainly in $\Delta t_{\rm jet}$. For the accretion mechanism \citep[e.g.][]{zhangdai10} and magnetic mechanism \citep[e.g.][]{kluzniak98,dai98,ruderman00}, $\Delta t_{\rm wait} \sim 0$ s. The jet launching time $\Delta t_{\rm jet}$ is mostly controlled by $\Delta t_{\rm clean}$, which may range from milliseconds (for most optimistic scenarios, e.g. \citealt{kluzniak98,ruderman00}) to $\sim 10$ s \citep{metzger11}. In any case, $\Delta t_{\rm bo}$ would contribute significantly to the total $\Delta t$. For the spindown model \citep{usov92,usov94}, the assumption is that a GRB jet is launched as the magnetar spins down. This would be after the early accretion phase. As a result, one should introduce an early $\Delta t_{\rm wait}$ for the duration of the accretion phase, which would be typically $1-10^3$ s. This is based on the observed duration of long GRBs (which are interpreted as the accretion time for most models) and an estimate of the free-fall timescale of a massive star $t_{\rm ff} \sim 180 \ {\rm s}$, which is the minimum timescale for accretion. As a result, this model would have a longer $\Delta t \sim (10-10^3)$ s than other magnetar models, with the delay mostly contributed from $\Delta t_{\rm wait}$.

\section{Case studies}\label{sec:cases}

\subsection{GW170817/GRB170817A association}\label{sec:170817}

This is the only robust GW-GRB association case. The GRB trigger time is delayed by $\sim 1.7$ s with respect to the binary NS merger time \citep{GW170817/GRB170817A}. The duration of GRB 170817A is $\sim 2$ s, with two pulses each lasting $\sim 1$ s  \citep{goldstein17,zhangbb18b}. Since this was a relatively weak burst, the true duration of each component ($t_{\rm pulse}$) could be longer than 1 s. 

According to the theory discussed above, there are two possible interpretations to the $\sim 1.7$ s delay. 

The first scenario invokes a matter/radiation-dominated fireball. Within this scenario, the sGRB is most likely emission from the photosphere. Using Eq.(\ref{eq:DeltatGRB}) and noticing $L \sim 10^{47} \ {\rm erg \ s^{-1}}$ for GRB 170817A \citep{goldstein17,zhangbb18b}, one has $\Delta t_{\rm GRB} \sim (20 {\rm ms}) \ L_{47} \Gamma_1^{-5}$. Forcing $\Delta t_{\rm GRB} \sim 2$ s, one requires very low Lorentz factor $\Gamma \sim 4$. This was the suggested ``cocoon breakout model'' (or mildly relativistic, wide angle outflow model) of GRB 170817A shortly after the discovery of the event \citep[e.g.][]{kasliwal17,mooley18a}, which was later disfavored by the discovery of superluminal motion of the radio afterglow of the source \citep{mooley18b,ghirlanda19}. Modeling of radio afterglow of the source suggests that the outflow Lorentz factor decays with time as 
$\Gamma \sim 4 (t/{\rm 10 \ day})^{-0.29}$ \citep{nakar18}, which means that at $t<10$ d, one has $\Gamma \gg 4$. This suggests that within this scenario, one has $\Delta t_{\rm GRB} \ll \Delta t$, and the delay should be attributed to $\Delta t_{\rm jet}$ and $\Delta t_{\rm bo}$. One way to make a long $\Delta t_{\rm jet}$ is to introduce a long $\Delta t_{\rm wait} \sim 1$ s, which is regarded as the duration of the HMNS phase. Such a waiting time was indeed introduced in some of the numerical simulations \citep[e.g.][]{gottlieb18,bromberg18}, and was suggested from kilonova modeling as well \citep{margalit17,gill19}. A long waiting time is also the necessary condition to make a significant cocoon emission component \citep{geng19}. This scenario has to assume that no relativistic jet is launched during the HMNS phase \citep{ruiz18,rezzolla18}, in contrast to some previous sGRB models \citep[e.g.][]{rosswog03}. The issue of this scenario is that one has to interpret the delay time $\Delta t \sim 1.7$ s and the duration of the burst $T_{90} \sim 2$ s using two different mechanisms: while $\Delta t$ is mostly controlled by $\Delta t_{\rm wait}$, $T_{90}$ has to be defined by the duration of the central engine (accretion timescale). The similar values of the two timescales have to be explained as a coincidence.

The second scenario, as advocated by \cite{zhangbb18b}, attributes $\Delta t$ mostly to $\Delta t_{\rm GRB}$. This is motivated by the intriguing fact that $\Delta t \sim 1.7$ s and $T_{90} \sim 2$ s are comparable. Based on Eq.(\ref{eq:DeltatGRB}), for a Poynting-flux-dominated outflow, $\Delta t_{\rm GRB} \sim t_{\rm pulse}$. If one takes the first pulse only and considers the weak nature of GRB 170817A (the true pulse duration should be longer than what is observed), one has $\Delta t_{\rm GRB} > 1$ s, which occupies most of the observed $\Delta t$. Within this scenario, both $\Delta t_{\rm jet}$ and $\Delta t_{\rm bo}$ are short (say, $\ll 0.5$ s), which suggests a negligible $\Delta t_{\rm wait}$. Within this scenario, there is no need to introduce an HMNS. The engine could be a BH, an HMNS with a lifetime shorter than 100 ms, an SMNS or even an SNS. One prediction of such a scenario is that $\Delta t$ should be correlated with the burst duration (if the bursts have 1-2 simple pulses like GRB 170817A). For example, if the next NS-NS-merger-associated sGRB has a shorter duration (e.g. 0.5 s), the delay time $\Delta t$ should be also correspondingly shorter. A smaller $\Delta t_{\rm wait}$ also suggests a less significant cocoon emission, even though the outflow is still a structured jet \citep{geng19}. 

\subsection{GW150914/GW150914-GBM association}

Since observationally the case is not robust, this association may not be physical. On the other hand, if the association is real, the delay timescale $\Delta t \sim 0.4$ s places great constraints on the proposed models. 

Most proposed models to interpret GW150914-GBM invoke substantial matter around the merging site. \cite{loeb16} invoked two BHs formed during the collapse of a massive star. Accretion after the merger powers the putative GRB. Putting aside other criticisms to the model \citep[e.g.][]{woosley16,dail17}, $\Delta t$ in such core collapse model should be at least $\Delta t_{\rm bo}$, which is 10s of seconds. The observed $\Delta t \sim 0.4$ s therefore essentially rules out the model, unless a contrived jet launching time is introduced \citep{dorazio18}. The same applies to other models that invoke a massive star as the progenitor of the putative GRB \citep[e.g.][]{janiuk17}. The reactivated accretion disk model \citep{perna16} is not subject to this constraint. However, it is likely that the reactivation happens way before the merger itself \citep{kimura17}.

The charged BH merger model (\cite{zhang16a}, see \cite{zhang19} for a more general discussion of charged CBC signals and \cite{dai19} for related signals) does not invoke a matter envelope surrounding the merger system. Within the framework discussed in this paper, both $\Delta t_{\rm jet}$ and $\Delta t_{\rm bo}$ are $\sim 0$ s, and $\Delta t$ is dominated by $\Delta t_{\rm GRB}$ (see \citealt{zhang16a} for detailed discussion). The difficulty of this model is the origin of the enormous charge needed to account for the GRB.

\section{Summary and discussion}\label{sec:summary}

We have discussed various physical processes that may cause a time delay $\Delta t$ of a GRB associated with a GW event. 
The conclusions can be summarized as follows:
\begin{itemize}
\item In general, there are three timescales, i.e. $\Delta t_{\rm jet}$, $\Delta t_{\rm bo}$ and $\Delta t_{\rm GRB}$, that will contribute to the observed $\Delta t$. Since the GRB jet launching mechanism is poorly understood, different scenarios make different assumptions. The assumptions introduced in different scenarios are sometimes contradictory (e.g. regarding whether a BH is needed to launch a GRB jet). The results are summarized in Table \ref{tab:Deltat}. With the GW information (which sets the fiducial time), one can in principle test these scenarios with a sample of GW/GRB associations in the future.
\item We considered both CBC-sGRB associations and GWB-lGRB associations. For the former, different models point towards a similar range of $\Delta t$: from 10 ms to a few seconds. The 1.7 s delay of GW170817/GRB 170817A falls into this range, so this duration alone cannot be used to diagnose the jet launching mechanism. On the other hand, a statistical sample of GW/GRB associations can in principle test the two scenarios with and without a significant intrinsic central engine waiting time: If $\Delta t$ of different events are independent of the duration of the sGRB and especially the duration of the GRB pulses $t_{\rm pulse}$, then it would be likely controlled by $\Delta t_{\rm jet}$, in particular, $\Delta t_{\rm wait}$ of the central engine. Such a $\Delta t_{\rm wait}$ may be attributed to the lifetime of an HMNS, and a BH engine is needed to power a sGRB. If, on the other hand, $\Delta t$ is always roughly proportional to $t_{\rm pulse}$, as is the case of GW170817/GRB 170817A association, then $\Delta t$ is likely dominated by $\Delta t_{\rm GRB}$, with negligible $\Delta t_{\rm jet}$ and $\Delta t_{\rm bo}$. The jet composition in this case is likely Poynting-flux-dominated, and the launch of a GRB jet may not necessarily require the formation of a BH.
\item The GWB-lGRB associations all should have a longer delay, with $\Delta t$ at least defined by the jet propagation and breakout time $\Delta t_{\rm bo}$, which is at least $\sim 10$ s. In some models, such as the Type II collapsar model and the magnetar spindown model, there could be an additional waiting time $\Delta t_{\rm wait}$ before the presumed jet launching mechanism starts to operate. In these cases, the delay can be as long as minutes to even hours. Detecting GWB-lGRB associations with such a long delay (e.g. $\gg 100$ s) would point towards these specific jet launching mechanisms or a progenitor star much larger in size than a Wolf-Rayet star.
\end{itemize}

\acknowledgments
{ I thank Wei-Hua Lei, Robert Mochkovitch, and Bin-Bin Zhang for discussion on the origin of $\Delta t$ of GW170817/GRB170817A association.}



\end{document}